\documentclass[doublecol]{epl2} 

\usepackage{amsmath,times}
\usepackage{amssymb,times}
\usepackage{amsthm,times}
\usepackage{enumerate,times}

\newtheorem{thm}{Theorem}

\newcommand{\dist}{\operatorname{dist}}

\renewcommand{\AA}{\mathcal A}

\newcommand{\EE}{\mathcal E}

\newcommand{\HH}{\mathcal H}

\newcommand{\KK}{\mathcal K}
\newcommand{\LL}{\mathcal L}

\newcommand{\R}{\mathbb R}

\newcommand{\Z}{\mathbb Z}

\title{Random Field Induced Order in Low Dimension}
\shorttitle{Random Field Induced} 

\author{Nicholas Crawford}
\shortauthor{N.~Crawford}

\institute{  The Technion, Haifa, Israel
}

\abstract{
We address an unresolved issue in the physics of low-dimensional many-body systems: the question of whether or not a random field can produce order at low temperatures for statistical mechanical systems possessing continuous internal symmetries.  
Concretely, we verify that the XY model in a uniaxial random field orders in two and three dimensions.  The direction the system orders is perpendicular to the randomness for \textit{any} choice of symmetry breaking field with nonzero projection perpendicular to the randomness.  The result is particularly relevant in two dimensions, where there are a number of competing effects - quasi-long range order of the pure system and strong fluctuations of the random field.   While we consider only classical systems explicitly, the effect is robust and our work has implications for quantum systems as well, producing ordered phases any dimension.
}

\begin{document}

\maketitle

\noindent
\textit{Introduction.}
It is generally understood that disorder plays an important role in the behavior of one-particle and many-body physics.   The most well-known example stems from the work of Anderson \cite{And} on transport properties of electrons in crystals with impurities; disorder can localize electron wave functions making conductivity negligible.   Other important examples of the effects of disorder include the "rounding" of first order transitions in low dimensional classical equilibrium systems \cite{AW, IM}.  This work was recently extended to quantum systems in \cite{ALR}.

The rounding effect aside, the behavior of interacting many-body systems in the presence of disorder is less well understood.  For example, it was only recently demonstrated by an in-depth analysis that Anderson localization persists in the presence of weak interactions between electrons \cite{AAB1}.  Localization of Cooper pairs was recently observed in the vicinity of the BCS superconducting transition due to the presence of disorder \cite{FermiLoc}.

In these examples, as well as in subjects such as quantum computing, the existence of quenched randomness in a system is viewed as an unwanted effect which disrupts ordering and coherence.  
It is less well known that randomness can itself \textit{create} ordering for systems in which the order parameter has a continuous symmetry.  We refer to this effect below as random field induced order (RFIO).
 
The basic model for RFIO is an XY model in a uniaxial random field. It is a classical equilibrium statistical mechanics model in which spins take values on the unit circle and interact in a ferromagnetic way.  There is additionally a random field acting in the vertical direction only.  
In more precise terms and generalizing the setting, we will below consider the following family of models.  Let $\Lambda_N=\left\{-N/2,\:...\:, N/2-1\right\}^d$.
For $x \in \Lambda_N$, spin variables $\sigma_x$ lie in the unit sphere $\mathbb S^{n-1} \subset \R^n$, $n \geq 2$.  We define the (random) Hamiltonian via
\begin{equation}
\label{E:Ham1}
-\HH_{N}(\sigma)= -\sum_{\langle x y\rangle} [\sigma_x- \sigma_y]^2 + \epsilon \sum_{x} \alpha_x \cdot \sigma_x + \sum_{x \in \partial \Lambda_N} u \cdot \sigma_x
\end{equation}
The first sum is over adjacent pairs $x, y$ in $\Lambda_N$ while the final sum consists of vertices at the boundary of $\Lambda_N$.  The $\alpha_x$'s form a family of independent, identically distributed  $k$-dimensional standard Gaussian vectors, where we have taken $k< n$.  The last term is for symmetry breaking purposes.
When defining Gibbs measures and states, the \textit{a priori} measure on spin space is uniform surface measure on $\mathbb S^{n-1}$.
We refer below to these models collectively as RFO$(n;k)$ models.  In this notation, the XY model with a uniaxial field is the RFO$(2;1)$ model.  The latter will be shortened to the RFO$(2)$ model.

The RFO$(2)$ model was first studied in the 1970's via mean field theory and a renormalization group calculation \cite{AA}.  The system appears to order, but only perpendicular to the axis on which the randomness acts.
In low dimensions, particularly on the two-dimensional square lattice $\Z^2$, the mean field and renormalization group calculations are less reliable, as the conclusions run counter to other effects, such as the Mermin-Wagner theorem and the Imry-Ma argument that strong fluctuations disrupt ordering \cite{IM}  (more about the relation between RFIO and these results appears below).  Thus in the 1980's multiple groups addressed two-dimensional behavior by more specialized renormalization group analyses \cite{DF1, DF2, MP}.  One group \cite{DF1} concluded there is a low temperature paramagnetic phase  while the other \cite{MP} concluded there is an intermediate-temperature ordered phase from which they extrapolate the low-temperature behavior.  Interesting, but only tangentially related, rigorous work was done in the 1990's on ground states in the strong field regime in \cite{Feld1, Feld2}.

Both analyses of the RFO(2) model in a weak field are problematic. The work \cite{DF1, DF2} is based on an approximation via a random field sine-Gordon model, which is extremely unstable with respect to the noise (see our related discussion of the random field Gaussian model below).  The paper \cite{MP} suggests the ordered phase persists at a fixed temperature even as the field strength $\epsilon$ is taken to $0$.  Such a conclusion is dubious as simpler versions of the model, such as a model which replaces the random field by a periodic field with period two has a critical inverse temperature which diverges as the field strength tends to $0$ in two dimensions.

With this background it seems useful to provide a definitive analysis of the the RFO$(2)$ model in two dimensions.  We do this and more in two main results below, the method of verification being a rigorous proof.
In Theorem \ref{T:Main1}, we state, for dimension $d= 2, 3$, that ordering in the direction perpendicular to randomness does occur at low temperature.  Moreover we show that this is the \textit{only} direction in which ordering occurs in a strong sense: it is independent of the direction of infinitesimal field used to break symmetry.  This is part of a more general result for RFO$(n;n-1)$ models.  A second result, Theorem \ref{T:Main2}, deals with RFO$(n;k)$ models for any $k< n$ and in any dimension $d\geq 2$:  The projection of the spin variable at any fixed vertex onto the subspace which supports the randomness is necessarily small.    

Let us remark that the issue of RFIO resurfaced over the last five years in the work of three separate groups.
With the advent of controllable interacting Bose-Einstein condensates in optical traps, the effect was suggested as a possible response to the presence of certain kinds of experimentally realizable disorder \cite{Wehr-et-al-2, SPL-Nature}.  Here the (pseudo-)spin variables arise from internal structure of the atoms, the tuning of interactions and the structure of the optical lattice.
Building on \cite{IK}, the same sort of mechanism was put forward as the reason for the splitting of Landau level degeneracy in experiments on graphene \cite{ALL}, though the particulars of this model, namely the fact that the derived random field is the curl of random coupling constants, means that the fluctuations of the disorder are subcritical from the perspective of the Imry-Ma argument. 
Finally, in \cite{vE3}, this question was posed during investigations of whether real-space coarse-graining procedures preserved a version of the spatial Markov property which characterizes Gibbs measures.

To explain the prior controversy and subtlety of our results in two dimensions we recall, for the convenience of the reader, the behavior of related models.

\noindent
\textit{The Pure O$(2)$ Model.}
In our language this is the case we take $n = 2$ and $\epsilon=0$, so there is no randomness.  When $d=2$ all Gibbs states are rotationally invariant in the thermodynamic limit (this is the content of the Mermin-Wagner theorem, see \cite{MW, DS} among many other works) and, in particular, there is no residual magnetic order.  There is however a Kosterlitz-Thouless phase transition \cite{FS, KT} expressed by a change in the behavior of the spin-spin correlation function
 $\langle \sigma_x \cdot \sigma_y\rangle$ from exponential to algebraic decay in terms of $|x-y|$.  If $d\geq 3$ residual magnetic ordering does occur \cite{FSS}.

\noindent
\textit{The Random Field Ising Model.}
In this case we constrain the spins to point only in the vertical direction, the same as the random field.
When $d \geq 3$ and weak disorder, it turns out that residual magnetic ordering persists \cite{Imbrie, BK}.  In dimension $d=2$ there is no magnetic order for any inverse temperature $\beta$ and any strength $\epsilon$, \cite{AW}.  This is the rounding effect alluded to in the introduction.  Physically, see \cite{IM}, the reason for this is that at all scales there are random fluctuations strong enough to overcome the loss in energy due to mismatched spins on the interior and exterior of domains.  

\noindent
\textit{The Random Field Gaussian Model.}
The last model we wish to mention is obtained by replacing the vector valued spins $\sigma_x \in \mathbb S^1$ by a field $\phi_x \in \R$ and otherwise retaining the setup described above.  This model appears, among other places, in \cite{vE0, SVBO}.  
In a finite volume $\Lambda$ with $0$ boundary conditions, we let $-\Delta_{\Lambda}$ denote the discrete Laplace operator on $\Lambda$ with Dirichlet boundary conditions.
We have
\begin{equation}
\mathbb E[\langle\phi_x \phi_y \rangle_{\Lambda}]= -\Delta_{\Lambda}^{-1}(x, y)+ \epsilon^2 \Delta_{\Lambda}^{-2}(x, y).
\end{equation}
Here and below $\mathbb E$ denotes the average over the randomness in the model of interest.

If $\epsilon > 0$ and if $d \leq 4$, the second term  on the RHS grows with $\Lambda$ while when $d=2$, it grows even after taking gradients in both arguments $x, y$.  

If one believes in magnetic ordering for the RFO$(2)$ model in two dimensions, the last example in particular should give pause as this is precisely what is obtained by expanding the RFO$(2)$ Hamiltonian \eqref{E:Ham1} in angular coordinates around a fixed direction.  
It suggests that either the ansatz of having order is flawed or that the fact that spins are constrained to lie on $\mathbb S^1$ plays a crucial role in keeping the influence of field fluctuations under control.
This effect is presumably the reason for the paramagnetic low temperature phase prediction in \cite{DF1}.  On the other hand it seems to have been neglected in the arguments presented in \cite{Wehr-et-al-2, MP}.  

\noindent
\textit{Order-by-Disorder.}
The phenomenon of "Order-by-Disorder" provides a class of systems which exhibit ordering due to various types of fluctuations.  The most relevant example, first considered by Henley \cite{H}, concerns a model Hamiltonian on $\Z^2$ of the form
\begin{equation*}
-\HH(\sigma )= \sum_{\|x-y\|_2=1} J_1 [\sigma_x- \sigma_y]^2 + \sum_{\|x-y\|_2^2} J_2 [\sigma_x- \sigma_y]^2
\end{equation*}
with $|J_1|< 2J_2$.  The  ground-states for this (frustrated) system are obtained by choosing a purely anti-aligned configuration of spins on each of the even and odd sub-lattices of $\Z^2$ and are thus parameterized by two angles:  an angle between the spin at $(0, 0)$ and the $e_1$-axis and relative angle between the spin at $(0, 1)$ and $(0, 0)$.  The degeneracy of ground-states is partially lifted under the  introduction of two types of "disorder".  The first type consists in passing from $0$ to positive temperature.  More relevant to the RFIO is a second mechanism: site dilution. Vertices of $\Z^2$ are deleted from the system independently with probability $p\ll 1$.  According to the calculations in \cite{H}, at $0$  and low temperature the system prefers the ground states with \textit{relative} angle between spins at $(0, 0)$ and $(0, 1)$ to be fixed at  $\pm \frac \pi2$.  

The crucial difference between this site diluted model and the RFO(n) models we consider concerns fluctuations due to randomness. These are substantially weaker in the site diluted order by-disorder setting.  In particular, there is an analog to the field $g_x$ introduced below but the fluctuations of this field are about as singular as the four dimensional version of the RFO(2) model.  While the site diluted order-by-disorder problem has not been rigorously addressed, this feature suggests the conclusions in \cite{H} are reliable. It would be interesting to see if our methods can be adapted to this case.  

\noindent
\textit{Main Results:}
In the following statements, $\texttt{P}$ denotes the projection operator onto the $k$ dimensional subspace supporting the distribution of $\alpha_x$.

Recall \eqref{E:Ham1} and let $\langle \cdot \rangle_{N}^{\beta, hu}$ denote the corresponding random Gibbs state on $\Lambda_N$ at inverse temperature $\beta$. For the statement of our main result, let us define the block average observables
$$
M_z= \epsilon^{d}\sum_{y:|y-z| \leq (2\epsilon)^{-1}} \sigma_y.
$$
\begin{thm}
\label{T:Main1} Let $d \in \{2, 3\}$ be fixed and  consider the RFO(n;n-1) model on $\Z^d$.  Let $e_n$ be a fixed unit vector so that $\texttt{P} \cdot e_n=0$.
There is an interval  $(0, \epsilon_0)$ so that the following holds:  
\vspace{5pt}

\noindent
For any $\epsilon \in(0, \epsilon_0)$ there are $\xi(\epsilon)> 0|\log \epsilon|^{-1/2}$ and
$\beta_0(\epsilon, \xi)>0$  so that if 
$$u \in \mathbb S^{n-1}, u \cdot e_n > \xi$$ 
and $\beta> \beta_0$, we have
$$
 \liminf_{N \uparrow \infty} \: \mathbb{E} \left[ \langle M_z\cdot e_n\rangle_N^{\beta, hu} \right]\geq 1- \xi
$$
for any $z\in \Z^d$.
Moreover $\xi(\epsilon) \rightarrow 0$ as $\epsilon \rightarrow 0$.
 \end{thm}
 \noindent
A similar effect was previously observed in XY models with a weak uniaxial field alternating direction in a chessboard fashion in classical \cite{AA, vE3} and quantum \cite{Aiz} models.

The condition $u \cdot e_n > \xi$ is technical; the result should hold for any choice $u \cdot e_n > 0$ independent of $\epsilon$ as long as $\epsilon$ is small.  It is possible that further analysis will remove this condition.  The proof of Theorem \ref{T:Main1} also gives lower bounds on $\beta_0(\epsilon)$ of order $\epsilon^{-2}$.  The analysis of reference \cite{MP} suggests the critical temperature is $\epsilon$-independent.  Resolving this discrepancy is an interesting open issue.

Even at $0$ temperature, Theorem \ref{T:Main1} strengthens the picture presented in \cite{Wehr-et-al-2, MP}.  Our methods show that, as $\epsilon$ tends to $0$, in the ground state the length of the projection of most spins onto the direction perpendicular to the randomness tends to $1$.

Intuition and \cite{AA} suggest similar statements should hold at a rigorous level for $d \geq 4$ and moreover that the effects of large fluctuations should be easier to control.  Our technical estimates bear this out, but currently we cannot provide a complete proof in this case (see the discussion below).  This is connected with the fact that control of the Dirichlet energy of functions provides better point-wise control in low dimension than in high dimension. 

Next we present a weaker result for general RFO$(n;k)$ models in any dimension $d\geq 2$.  Let $\sigma_0$ be a fixed spin configuration on $\Z^d$ and let $\langle \cdot \rangle_N^{\sigma_0}$ denote the Gibbs state on $\Lambda_N$ with boundary condition $\sigma_0$
\begin{thm}
\label{T:Main2} Let $d \geq 2$ and $k < n$ be fixed and  consider the RFO$(n;k)$ model on $\Z^d$.
There is an interval  $(0, \epsilon_0)$ so that for any boundary condition $\sigma_0$ the following holds:
\vspace{5pt}

\noindent
For any $\epsilon \in(0, \epsilon_0)$  there exists a
$\beta_0(\epsilon)$ so that for the RFO$(n; k)$ model with$\epsilon$and $\beta> \beta_0$  
$$
\limsup_N \:  \mathbb{E} \left[ \langle \| \texttt{P} \cdot M_z\|_2^2\rangle^{\beta, \sigma_0}_N \right]\leq \epsilon.
$$
 \end{thm}
 \noindent
For $d \in\{2, 3, 4\}$, Theorem \ref{T:Main2}  is close in spirit to Theorem 4.4 in the paper by Aizenman and Wehr \cite{AW}, though neither result implies the other.  

We make no rigorous statements regarding magnetic ordering in RFO$(n; k)$ models for $k \leq n-2$, although ordering should occur if $d\geq3$.  If $d=2$ we expect the RFO$(n;n-2)$ models have a Kosterlitz-Thouless transition and that if $k< n-2$, the model exhibits exponential decay of correlations for all $\beta$ in correspondence with the expected behavior of pure O$(n-k)$ models when $d=2$.

The choice of standard Gaussian vectors is not crucial.
The proof of Theorem \ref{T:Main1} applies for variables with sub-Gaussian tails in $d=2$ and for any distribution having arbitrary exponential moments if $d=3$. The proof of Theorem \ref{T:Main2} requires even weaker assumptions on the tail behavior.


\noindent
\textit{Extensions:}
The effect described in this note should be very robust, applying to classical and quantum systems and both symmetric and biased disorder.
Examples include:
\vspace{2pt}

\noindent \texttt{Biased Disorder:}
Suppose the $\vec{\alpha}_x=\sum_{i=1}^m \vec{\alpha}_x(i)$ with $\vec{\alpha}_x(i)$ independent Gaussian vectors with non-zero means $\mathbb E[\vec{\alpha}_x(i)]= v_i$.  For example, if the $v_i$ from an orthonormal basis for $\R^n$ ordering will occur along the directions $\sum_{i} \pm v_i$. Such a model should be treatable by our methods combined with Pirogov-Sinai theory.

\noindent
\texttt{Quantum models:}  Whenever the quantum system without disorder  may be expressed in terms of pseudo-spin order parameters with continuous symmetry, effects analogous to those described above are possible. Examples include 
Bose and Fermi Hubbard models as discussed in \cite{SPL-Nature} and graphene with fluctuating interaction strengths \cite{ALL}.  

\noindent
\textit{Proofs.}
For the interested reader, we now give a brief sketch of the proof.
For simplicity, we restrict the discussion to the RFO$(2)$ model. Demonstration of the remainder of Theorem \ref{T:Main1} follows along the same lines.  The proof of Theorem \ref{T:Main2} is actually simpler because we do not need to distinguish different phases.
\label{S:Intuit}
For notational purposes set
\begin{equation}
\epsilon_d=
\begin{cases}
\epsilon \sqrt{|\log \epsilon|} \quad \text{ if $d=2$},\\
\epsilon \quad \text{ if $d \geq 3$}.
\end{cases}
\end{equation}
We use the notation $|R|$ for the cardinality of a finite set $R\subset \Z^d$.

\noindent
 \textbf{Calculations for Small Boxes.} 
To understand the issues we encountered in the analysis, let us begin by studying the model restricted to boxes  $Q_{\ell} \subset \Lambda$.  We take the side-length $\ell$ of order $\ell \sim \epsilon_d^{-1}$
 without being completely precise yet.
 This scale is the fundamental length scale in the problem as revealed below in formula 
 \eqref{E:COV}.  
   
In a fixed box $Q_{\ell}$, the first calculation we do is an optimization of the Hamiltonian $-\HH_{Q_\ell}$ assuming the spin variables have small deviations from a fixed direction.  Here $-\HH_{Q_\ell}$ is the analog of \eqref{E:Ham1} in $Q_\ell$ with free boundary conditions.   

For convenience, denote $\hat{\alpha}_x:= \alpha_x - |Q_{\ell}|^{-1} \sum_{z\in Q_{\ell}} \alpha_z$ and 
$$
\text{E}_{Q_\ell}(\alpha):= -\sum_{x \in Q_{\ell}} \hat{\alpha}_x \cdot \Delta^{-1} \hat{\alpha}_x.
$$

Expanding $-\HH_{Q_\ell}$ to second order in the $\hat{\theta}$ variables,
\begin{equation}
\label{E:Opt}
\sup_{\stackrel{(\theta_x)_{x \in Q_{\ell}}}{\theta_x \approx \psi}} -\HH_{Q_\ell}(\theta)
= \frac{\epsilon^2}{2}\cos^2(\psi)\text{E}_{Q_\ell}(\alpha) + O\left(\epsilon\sum_{z \in Q_{\ell}} \alpha_z\right).
\end{equation}
The first term has the typical order of magnitude $\epsilon^2 \ell^d$ for $d\geq 3$ and $\epsilon^2 \ell^2 \log \ell$ for $d=2$ while the second term is typically of order $\epsilon \ell^{d/2}$.   

Neglecting the second term, directions of presumed ordering are obtained by neglecting the second term and optimizing the RHS in $\psi$.
For $\psi$ fixed, the optimal choice for the deviation variables $\hat{\theta}_x$ is $\hat{\theta}_x=  \epsilon \cos(\psi) g_{Q_{\ell},x}$ where
$g_{Q_{\ell}, x}= -\Delta^{-1} \cdot \hat{\alpha}_x$.

In evaluating the validity of this calculation one has to worry about three issues.  
First, we must understand in what sense the assumption of small deviation from a fixed angle is valid.  
If we denote the Dirichlet energy of spin configurations in boxes $Q_\ell$ by
$$
\EE_{Q_\ell}(\sigma):= \frac {1}{2}\sum_{\stackrel{|x- y|=1}{x, y \in Q_{\ell}}} [\sigma_x- \sigma_y]^2,
$$
the only \textit{ a priori} control of small amplitude deviations we are able to obtain regarding $\EE_{Q_\ell}$
is that it is costly energetically  for
$\EE_{Q_\ell}(\sigma) \geq 4 \epsilon^2_d \: |Q_{\ell}|.$
 One cannot ask for more than this because approximate ground states, which take the form $
 \sigma_x= (\cos(\epsilon g_{Q_{\ell}, x}), \sin(\epsilon g_{Q_{\ell}, x}))$, have this Dirichlet energy.

Second, the calculation itself imposes an upper bound on $\ell$:  It is a fact that $g_{Q_\ell, x}$ has typical order of magnitude in dimensions $d=2, 3, 4$ of $\ell, \sqrt{\ell}, \sqrt{\log \ell}$ respectively.  This indicates that naive expansions breakdown beyond the length scales $\epsilon^{-1},\: 
\epsilon^{-2}, \:
\exp{(\epsilon^{-2})}$, respectively since $\epsilon g$ is of constant order.  If we take fluctuations into account, in two dimensions this imposes $\ell \lesssim \epsilon_2^{-1}$.
If one believes in ordering, nonlinear effects must play a role at larger scales.

Conversely, and of particular significance in two dimensions, moderate deviations of the field may make the term $O\left(\epsilon\sum_{z \in Q_{\ell}} \alpha_z\right)$ relevant.  In two dimensions, this imposing a lower bound on $\ell \gtrsim \epsilon^{-1}|\log \epsilon|^{-1}$.  There is just enough room between the two constraints to make sense of the computation.  If we take $\ell = \epsilon^{-1}|\log \epsilon|^{-\frac 12-\gamma}$ for $\gamma \in (0, 1/2)$ fixed,  we are able to keep $|g_x| \leq |\log \epsilon|^{-\gamma}$ (even taking into account fluctuations) while 
$O\left(\epsilon\sum_{z \in Q_{\ell}} \alpha_z\right)$ is lower order with probability exponential in $|\log \epsilon|^{1-2\gamma}$.

\noindent
\textbf{A Key Computation.}
Next we present a computation of both physical and technical interest which allows us to turn the naive analysis presented above into a rigorous Peierls argument.  The idea is that low energy spin configurations are composed of a fast oscillating  approximate ground state (like $g_{Q_{\ell}, x}$ above) and a long wavelength contribution.  Subtracting off the fast oscillating component renormalizes the Hamiltonian into a more tractable form.

For any connected set $R \subset \Z^d$ (not necessarily a box), we introduce the Dirichlet Laplacian $\Delta_{R}^D$ and represent spin configurations on $R$ in angular coordinates by $\sigma_x= (\cos(\theta_x), \sin(\theta_x))$.  Using the field
$g'_x=[-\Delta_R^D+ \ell^{-2}]^{-1} \alpha_x$,
we make the change of variables $\phi_x = \theta_x- \epsilon \cos(\theta_x) g'_x$.
This change of variables should be compared with the optimizer for fixed angle $\psi$ described after \eqref{E:Opt}.
The mass$^2$ term $\ell^{-2}$ is introduced in order to keep the $g'_x$ small in modulus, as long as the field $\alpha_x$ behaves in a typical way inside the region $R$.

The Hamiltonian \eqref{E:Ham1} restricted to $R$ transforms according to $-\HH_{R}(\sigma|\sigma^0) \mapsto \KK(\phi_x|\phi^0)$
with
\begin{equation}
\label{E:COV}
\KK(\phi_x|\phi^0)= \sum_{\langle xy\rangle} [\cos(\phi_x- \phi_y)-1] + \frac 12 \sum_{x} m^2_x \cos^2(\phi_x)
\end{equation}
where $m^2_x= \sum_{y: |x-y|=1} [g'_y-g'_x]^2$.
There are errors in making this transformation, but these errors can be controlled if Dirichlet energy of $\sigma$ in $R$ is smaller than $4\epsilon_d^2 |R|$ and the field $\alpha$ behaves in a typical way inside $R$.
Note that this transformation introduces the length scale $\epsilon^{-1}_d$ in an explicit way - $m_x$ is typically of order $\epsilon_d$.  Thus, phase boundaries separating the two pure phases $\theta \sim 0$ and $\theta \sim \pi$ have width of order $\epsilon_d^{-1}$.

\noindent
\textbf{The Modified Peierls Argument.}
Next we explain how to use \eqref{E:COV} to implement a Peierls argument.
Taking inspiration from  \cite{Pres-Book},
we define a second scale 
$$
L \sim
\begin{cases}
\epsilon^{-1} \log^4 \epsilon \quad \text{ if $d=3$},\\
\epsilon^{-1} |\log \epsilon|^{-\frac 12+ \gamma} \quad \text{ if $d=2$},
\end{cases}
$$ 
so that $\ell\ll \epsilon_d^{-1}\ll L$. 

Given a spin configuration $\sigma$ on $\Lambda$, a box $Q_{\ell}\subset \Gamma$ is bad for $\sigma$ if one of two things occurs.  THe first possibility is that the Dirichlet energy in $Q_\ell$ is substantially larger than the energy scale $\text{E}_{Q_\ell}(\alpha)$ set by the randomness in $Q_{\ell}$  (this is typically $O(\epsilon_d^2 \ell^d$).  The second possible source of bad behavior, if the Dirichlet energy is smaller than this scale, is that the average angle $\psi_{Q_{\ell}}$ associated with $\sigma$ in $Q_\ell$ is bounded away from  $\{0, \pi\}$ by a chosen cutoff $\xi$.  

A region of space $\Gamma \subset \Lambda$ is called \textit{contour} for $\sigma$ if it is maximally connected union of boxes $\{Q_L(z): z \in L\Z^d\}$ so that for each $Q_{L}$, there is a bad cube $Q_\ell$ within distance $3L/2$.  The goal, as with any Peierls argument, is to show that large contours are unlikely to occur at low temperature.

Given a  spin configuration $\sigma$ and an associated  contour $\Gamma$, to extract energy cost from the existence of $\Gamma$ we compare $\sigma$ with a new spin configuration $\tilde{\sigma}$ which agrees with either $\sigma$ or the reflection of $\sigma$ across the $e_2$ axis on each component of $\Lambda_N \backslash \Gamma$.  We require that $\tilde \sigma$ is within some $\delta\ll \xi$ of either $0$ or $\pi$ on the whole of $\Gamma$.  

We construct $\tilde \sigma$ in a few steps.   First, we find a layer $\LL$ surrounding $\Gamma$ which itself has "thickness" of order $L$. $\LL$ has two properties: at the boundary of $\LL$,  $|\sigma_x \cdot e_1|> 1/2$ for all $x$ and also $\sigma_x \cdot e_1$ is of constant sign on each connected component of $\LL $.    In what follows, a contour will be called a $\pm$ contour depending on the sign of $\sigma_x \cdot e_1$ on the component of $\LL$ which separates $\Gamma$ from $\infty$.

It is worth remarking that it is here that our restriction on the dimension enters.  By definition, at the boundary of a contour, the Dirichlet energy $\EE_{Q_\ell}$ is under control as is the spin average, $\ell^{-d} \sum_{x \in Q_{\ell}} \sigma_x$, for all boxes $Q_\ell \subset \LL$.  The lower the dimension, the more strongly this control restricts the size of "defects".
  
Having found the boundary layer $\LL$, we change to the $\phi$ variables inside this boundary layer, fixing the spins in $\Lambda_N \backslash  \LL$.
 This will cost us energetically, but less than we ultimately gain by extracting energy from the bad behavior on the contour.  

Next, we replace $\sigma$ in $\LL $ by the spin configuration $\sigma'$ associated with the optimizer of \eqref{E:COV} subject to the boundary condition given by $\sigma$ on $\Lambda_N \backslash \LL $.  In each connected component of $\LL$, one can show that, for typical realizations of disorder, $\sigma'$ is uniformly close to one of $\pm e_1$ as long as $\dist(x, \LL ^c)> L/4$.  This follows from the form of \eqref{E:COV} as long as $L\gg\epsilon_d^{-1}$. 

The final step is to modify $\sigma'$ as follows.  We first optimize $\KK_{\AA}(\phi)$ in the domian
$$
\AA:= \{ x \in \LL\cup \Gamma: \dist(x, \Lambda_N \backslash(\LL\cup \Gamma))> L/2\}
$$ 
with \textit{free} boundary conditions.  Typically the optimizer $\phi_{\AA}$ is close to $0$ throughout $\AA$.  We want to "glue" $\sigma'$ in $\AA^c$ to the spin configuration $\eta$  determined by $\phi_{\AA}$ with minimal cost in energy.  To do this, on each \textit{internal} component of $\Lambda_N \backslash \AA$ we replace, as necessary,  $\sigma'$ by its reflection $\sigma''$ across the $e_2$ axis so that $\sigma''_x \cdot e_1$ has the same sign on all components of $\LL \backslash \AA$.  $\tilde{\sigma}$ is defined as $\sigma''$ on $\Lambda_N \backslash \AA$ and is $\eta$ (or its reflection) on $\AA$.

The big hurdle in all of these considerations is to make sure regions where fluctuations of the disorder invalidate the above reasoning are sparse.  This is rather delicate in two dimensions.
Details may be found in \cite{BC}.

\noindent
\textit{Conclusion.}
In this letter we addressed a 25 year old controversy regarding randomness induced ordering, showing rigorously that the random field can actual select ordered phases in two and three dimensions.  The results demonstrate that the low dimensional qualitative behavior of such models agrees with the first work on the subject \cite{AA} and elucidates contradictory results in the papers \cite{DF1, MP}. 

There are two ways to understand these results in two dimensions.  On the one hand, if one thinks along the lines of Anderson localization, the randomness localizes the spin wave and vortex dipole excitations which enable the Kosterlitz-Thouless phase.  On the other hand, if one thinks about the homogenization theory of classical fields, the behavior we discuss is a consequence of imposing the hard constraint that spins lie on the unit sphere.

Our work gives a very precise picture for the behavior of these systems at low temperature, bounds on transition temperatures in terms of the strength of the randomness and tools to extend the stated results to a broad class of classical and quantum systems.

\end{document}